\documentclass[12pt]{scrartcl}

\usepackage{amsmath,amssymb}

\usepackage{graphicx}

\DeclareGraphicsExtensions{.eps,.pdf,.jpg,.png}

\begin{document}

\LARGE\bf\sffamily 
\begin{center}
Tunneling in
Fractional\\ Quantum Mechanics
\end{center}

\normalsize

\begin{center}
E. Capelas de Oliveira\footnote{capelas@ime.unicamp.br} 
and Jayme Vaz~Jr.\footnote{vaz@ime.unicamp.br} 
\end{center}

\normalsize\rm\sffamily

\begin{center}
Departamento de Matem\'atica Aplicada - IMECC\\
Universidade Estadual de Campinas \\
13083-859 Campinas, SP, Brazil
\end{center}

\bigskip

\small\rm\rmfamily

\begin{center}
\begin{minipage}[c]{10cm}
\centerline{\bf\sffamily Abstract}
We study the tunneling  through delta and 
double delta potentials in fractional quantum 
mechanics. After solving the fractional Schr\"odinger 
equation for these potentials, we calculate the 
corresponding reflection and transmission coefficients. 
These coefficients have a very interesting behaviour. 
In particular,  we can have zero 
energy tunneling when the order of the Riesz 
fractional derivative is different from 2. 
For both potentials, the zero energy limit 
of the transmission coefficient 
is given by $\mathcal{T}_0 = \cos^2{(\pi/\alpha)}$, 
where $\alpha$ is the order of the
derivative ($1 < \alpha \leq 2$). 
\end{minipage}
\end{center}

\normalsize\rm\rmfamily

\medskip

\section{Introduction}

In recent years the study of fractional integrodifferential equations 
applied to physics and other areas has grown. Some examples 
are \cite{Rang, Mainardi, Duan}, among many others. 
More recently, the fractional generalized Langevin equation 
is proposed to discuss the anomalous diffusive behavior of a 
harmonic oscillator driven by a two-parameter Mittag-Leffler 
noise \cite{Rubens}. 

Fractional Quantum Mechanics (FQM) is the theory of quantum 
mechanics based on the fractional Schr\"odinger equation (FSE). 
In this paper we consider the FSE as 
introduced by Laskin in \cite{Laskin1,Laskin2}. 
It was obtained in the context of the path integral approach to quantum mechanics. 
In this approach, path integrals are defined over L\'evy flight paths, which is a 
natural generalization of the Brownian motion \cite{Levy}. 

There are some papers in the literature studying solutions of FSE. 
Some examples are \cite{Ex1,Ex2,Naber}. However, recently Jeng et al. \cite{Jeng} 
have shown that some claims to solve the FSE have not taken into 
account the fact that the fractional derivation is a {\it nonlocal} operation. 
As a consequence, all those attempts based on local approaches are
intrinsically wrong. 
Jeng et al.\! pointed out that the only correct one 
they found is the one \cite{Dong} involving the delta potential. 
However, in \cite{Dong} the FSE with delta potential was 
studied only in the case of {\it negative} energies. 
This has been generalized in \cite{EFJ}, 
where we have solved the FSE for the delta and double 
delta potentials for positive and negative energies. 

The objective of this paper is to study the tunneling 
through delta and double delta potentials in the 
context of the FSE. As a result, we found some 
very interesting properties that are not observed 
in the usual $\alpha=2$ quantum mechanics. 
Probably the most interesting is the presence of tunneling 
through delta and double delta potentials  
even at zero energy. Moreover, in the case 
of the double delta potential, this zero energy 
tunneling is independent of the relation of the
two delta functions. In Lin et al. \cite{Lin} the 
problem of calculating the 
transmission coefficient in FQM for the double delta potential
has been addressed; however, the authors have 
used that same \textit{local} approach that Jeng et al. \cite{Jeng} 
have shown to be wrong. As expected in this case, our 
results differs from theirs. 

We organized this paper as follows. Firstly, and for the 
sake of completeness, we reproduce the solution of the FSE 
for the delta and double delta potentials, as given in \cite{EFJ}, 
presenting their 
respective solutions in terms of Fox's $H$-function. 
Some calculations and properties of the Fox's $H$-function are
given in the Appendixes. Then we study the asymptotic behaviour of
those solutions, calculate the reflection and 
transmission coefficients, and study some of 
their properties. The limit $\alpha \rightarrow 2$ 
for these coefficients and
the boundary conditions satisfied by the solutions
are also discussed in two other Appendixes.

\section{The Fractional Schr\"odinger Equation}

The one-dimensional FSE is 
\begin{equation}
\label{FSE_with_T}
i\hbar \frac{\partial\psi(x,t)}{\partial t} = D_\alpha(-\hbar^2 \triangle)^{\alpha/2}\psi(x,t) + V(x)\psi(x,t) , 
\end{equation}
where $1 < \alpha \leq 2$, $D_\alpha$ is a constant, $\triangle = \partial_x^2$ is the Laplacian, and $(-\hbar^2 \triangle)^{\alpha/2}$ 
is the Riesz fractional derivative \cite{Butzer}, that is, 
\begin{equation}
\label{Riesz}
(-\hbar^2 \triangle)^{\alpha/2} \psi(x,t) = \frac{1}{2\pi \hbar}\int_{-\infty}^{+\infty} 
{\mbox e}^{i p x/\hbar} | p |^\alpha \phi(p,t) {\mbox d}p , 
\end{equation}
where $\phi(p,t)$ is the Fourier transform of the wave function, 
\begin{equation}
\label{fourier}
\phi(p,t) = \int_{-\infty}^{+\infty} {\mbox e}^{-ipx/\hbar}\psi(x,t) {\mbox d}x , \qquad 
\psi(x,t) = \frac{1}{2\pi \hbar}\int_{-\infty}^{+\infty} {\mbox e}^{ipx/\hbar}\phi(p,t) {\mbox d}p . 
\end{equation}
The time-independent FSE is 
\begin{equation}
\label{FSE}
D_\alpha(-\hbar^2 \triangle)^{\alpha/2}\psi(x) + V(x)\psi(x) = 
E\psi(x) . 
\end{equation}
In the momentum representation, this equation is written as 
\begin{equation}
\label{MFSE}
D_\alpha |p|^\alpha \phi(p) + \frac{(W\ast \phi)(p)}{2\pi \hbar} = E \phi(p) , 
\end{equation}
where $(W\ast \phi)(p)$ is the convolution 
\begin{equation}
(W\ast \phi)(p) = \int_{-\infty}^{+\infty}W(p-q)\phi(q){\mbox d}q , 
\end{equation}
and $W(p) = \mathcal{F}[V(x)]$ is the Fourier transform of the
potential $V(x)$. 

Solutions of the FSE for delta and double delta potentials 
are given in \cite{EFJ} in the situations of bound and 
scattering states. Since we need to study the asymptotic 
behaviour of these solutions in order to find the 
transmission coefficients, and for the
sake of completeness, we will reproduce the 
calculations of the wave functions in the case of scattering states.

\subsection{FSE for Delta Potential}

Let us consider the case 
\begin{equation}
V(x) = V_0 \delta(x) , 
\end{equation}
where $\delta(x)$ is the Dirac delta function and $V_0$ is a constant. Its Fourier transform is 
$W(p) = V_0$ and the convolution $(W\ast \phi)(p)$ is 
\begin{equation}
(W\ast \phi)(p) = V_0 K , 
\end{equation}
where the constant $K$ is 
\begin{equation}
\label{eq_K}
K = \int_{-\infty}^{+\infty}\phi(q) {\mbox d}q . 
\end{equation}
The FSE in the momentum representation (\ref{MFSE}) is 
\begin{equation}
\label{MFSE_delta}
\left( |p|^\alpha - \frac{E}{D_\alpha}\right)\phi(p) = -\gamma K , 
\end{equation}
where 
\begin{equation}
\label{def_gamma}
\gamma = \frac{V_0}{2\pi \hbar D_\alpha} . 
\end{equation}

Since we are interested in scattering states, we 
will consider that $E > 0$ and write
\begin{equation}
\label{def_lamb_2}
\frac{E}{D_\alpha} = \lambda^\alpha , 
\end{equation}
where $\lambda > 0$. Since $f(x)\delta(x) = f(0)\delta(x)$, 
the solution of Eq.(\ref{MFSE_delta}) in this case is 
\begin{equation}
\label{sol_plus_1}
\phi(p) = \frac{-\gamma K}{|p|^\alpha - \lambda^\alpha} + 2\pi\hbar C_1 \delta(p-\lambda) + 2\pi\hbar C_2 \delta(p+\lambda) , 
\end{equation}
where $C_1$ and $C_2$ are arbitrary constants 
and the constant $2\pi\hbar$ was introduced for later convenience.
 Using this in Eq.(\ref{eq_K}) gives that 
\begin{equation}
\label{cond_K_2}
K = -\gamma K \int_{-\infty}^{+\infty} \frac{{\mbox d}p}{|p|^\alpha -\lambda^\alpha} + 
2\pi\hbar C_1 + 2\pi\hbar C_2 , 
\end{equation}
where the integral is interpreted in the sense of Cauchy principal value, and it gives 
\begin{equation}
\label{int2}
\int_{-\infty}^{+\infty} \frac{{\mbox d}p}{|p|^\alpha -\lambda^\alpha} = 2\lambda^{1-\alpha} 
\int_0^{+\infty}\frac{{\mbox d}q}{q^\alpha -1} = -2\lambda^{1-\alpha} \frac{\pi}{\alpha}\cot{\frac{\pi}{\alpha}} , 
\end{equation}
where we have used formula 3.241.3 (pg. 322) of \cite{Grad} - see Eq.(\ref{grad3.241.3}). The constant $K$ is therefore 
\begin{equation}
K = \frac{2\pi\hbar (C_1 + C_2)\alpha \lambda^{\alpha-1}}{\alpha\lambda^{\alpha-1}-
2\pi\gamma \cot{(\pi/\alpha)}} , 
\end{equation}
and we have 
\begin{equation}
\label{phi_p_plus} 
\phi(p) = 2\pi\hbar C_1\delta(p-\lambda) + 2\pi\hbar C_2\delta(p+\lambda) - 
\frac{2\pi\hbar\gamma (C_1 + C_2)\alpha \lambda^{\alpha-1}}{(\alpha\lambda^{\alpha-1}-2\pi\gamma \cot{(\pi/\alpha)})} 
\frac{1}{(|p|^\alpha-\lambda^\alpha)} . 
\end{equation}
Next we need to calculate the inverse Fourier transform of $\phi(p)$ to obtain $\psi(x)$, that is, 
\begin{equation}
\label{fourier.inv.delta.2}
\psi(x) = C_1 {\mbox e}^{i\lambda x/\hbar} + C_2 {\mbox e}^{-i\lambda x/\hbar} 
-\frac{2\pi\gamma\alpha (C_1 + C_2)}{(\alpha\lambda^{\alpha-1}-2\pi\gamma \cot{(\pi/\alpha)})} 
\mathfrak{J}_\alpha\left(\frac{\lambda x}{\hbar}\right) , 
\end{equation}
where $\mathfrak{J}_\alpha(w)$ is the Cauchy principal value of the integral 
\begin{equation}
\label{jota.alpha}
\mathfrak{J}_\alpha(w) = \frac{1}{\pi}\int_0^{+\infty} \frac{\cos{w q}}{q^\alpha -1}{\mbox d}q , 
\end{equation}
and such that 
\begin{equation}
\label{jota.alpha.aux}
\int_{-\infty}^{+\infty}\frac{{\mbox e}^{ipx/\hbar}}{|p|^\alpha-\lambda^\alpha} {\mbox d}p = 
2 \pi \lambda^{1-\alpha}\mathfrak{J}_\alpha(\lambda x/\hbar) . 
\end{equation}
The above integral is calculated in the Appendix B, 
and  is given by Eq.(\ref{JFox}). Using this result, and the definition of $\gamma$ in Eq.(\ref{def_gamma}) 
and $\lambda$ in Eq.(\ref{def_lamb_2}),  
we can write that 
\begin{equation}
\label{psi.delta.plus}
\psi(x) =  C_1 {\mbox e}^{i\lambda x/\hbar} + C_2 {\mbox e}^{-i\lambda x/\hbar} 
+ \Omega_\alpha \frac{ (C_1+C_2)}{2} \Phi_\alpha\left(\frac{\lambda |x|}{\hbar}\right) , 
\end{equation}
where 
\begin{equation}
\label{def_Phi}
\begin{split}
\Phi_\alpha\left(\frac{\lambda |x|}{\hbar}\right) & = 
\frac{\alpha \hbar}{\lambda |x|} \bigg(
 H_{2,3}^{2,1} \left[ \left( \frac{\lambda |x|}{\hbar}\right)^\alpha  \left| 
\begin{array}{l}
(1,1), (1,(2+\alpha)/2) \\
(1,\alpha), (1,1), (1,(2+\alpha)/2) \end{array} \right. \right] \\
& \quad - H_{2,3}^{2,1} \left[ \left( \frac{\lambda |x|}{\hbar}\right)^\alpha   \left| 
\begin{array}{l}
(1,1), (1,(2-\alpha)/2) \\
(1,\alpha), (1,1), (1,(2-\alpha)/2) \end{array} \right. \right] \bigg) ,  
\end{split}
\end{equation}
with  $H_{2,3}^{2,1} \left[ \cdot \left| 
\begin{array}{l}
- \\
- \end{array} \right. \right]$ denoting a Fox's $H$-function (see Appendix A), 
and 
\begin{equation}
\label{def_omega}
\Omega_\alpha = \bigg[\bigg(\frac{E}{U}\bigg)^{\frac{\alpha-1}{\alpha}}-\cot{\frac{\pi}{\alpha}}\bigg]^{-1} , 
\end{equation}
and 
\begin{equation}
\label{def_U}
U = \bigg(\frac{V_0}{\alpha \hbar D_\alpha^{1/\alpha}}\bigg)^{\alpha/(\alpha-1)} . 
\end{equation}

\subsection{FSE for Double Delta Potential}

Now let the potential be given by 
\begin{equation}
\label{delta.potential}
V(x) = V_0 [\delta(x+R/2) + \mu \delta(x-R/2)] , 
\end{equation}
with $\mu$, $V_0$ and $R$ real constants. 
When $V_0 < 0$ this potential can be seen as a 
model for the one-dimensional limit of the
molecular ion $H_2^+$ \cite{Scott}. The parameter
$R$ is interpreted as the internuclear distance and
the coupling parameters are $V_0$ and $\mu V_0$. 
Its Fourier transform is 
\begin{equation}
W(p) = V_0 {\mbox e}^{ipR/2\hbar} + V_0 \mu {\mbox e}^{-ipR/2\hbar} 
\end{equation}
and for the convolution 
\begin{equation}
(W\ast \phi)(p) = V_0 {\mbox e}^{ip R/2\hbar} K_1(R) + V_0 \mu {\mbox e}^{-ip R/2\hbar} K_2(R) , 
\end{equation}
where $K_1(R)$ and $K_2(R)$ are constants given by 
\begin{equation}
\label{Ks}
K_1(R) = K_2(-R) = \int_{-\infty}^{+\infty} {\mbox e}^{-i Rq/2\hbar} \phi(q) {\mbox d}q . 
\end{equation}
The FSE in momentum space is  
\begin{equation}
\label{MFSE_DD}
\left(|p|^\alpha - \frac{E}{D_\alpha}\right)\phi(p) = -\gamma {\mbox e}^{i R p/2\hbar} K_1(R) 
-\gamma \mu {\mbox e}^{-i R p/2\hbar} K_2(R) , 
\end{equation}
where we used the notation introduced in Eq.(\ref{def_gamma}). 

Since we are interested in scattering states, we have $E > 0$ and 
we write $\lambda$ as in Eq.(\ref{def_lamb_2}) 
and for the solution of Eq.(\ref{MFSE_DD}) we have 
\begin{equation}
\label{phi_dd_plus}
\phi(p) = 2\pi\hbar C_1 \delta(p-\lambda) + 2\pi\hbar C_2\delta(p+\lambda) -\frac{\gamma {\mbox e}^{i R p/2\hbar} K_1(R)}{|p|^\alpha-\lambda^\alpha} 
- \frac{\mu\gamma {\mbox e}^{-i R p/2\hbar} K_2(R)}{|p|^\alpha - \lambda^\alpha} .  
\end{equation} 
Using this expression for $\phi(p)$ in Eq.(\ref{Ks}) of definition of $K_1(R)$ and $K_2(R)$ we have 
\begin{gather}
(1+2\pi \gamma\lambda^{1-\alpha} \mathfrak{J}_\alpha(0))K_1(R) + \mu 2\pi\gamma\lambda^{1-\alpha} \mathfrak{J}_\alpha(\lambda R/\hbar) K_2(R) = 2\pi \hbar C^\prime_1 , \\
2\pi \gamma\lambda^{1-\alpha}\mathfrak{J}_\alpha(\lambda R/\hbar) K_1(R) + (1+\mu 2\pi\gamma\lambda^{1-\alpha} \mathfrak{J}_\alpha(0))K_2(R) = 2\pi \hbar C^\prime_2 , 
\end{gather}
where 
\begin{equation}
C^\prime_1 = C_1 {\mbox e}^{-iR\lambda/2\hbar} + C_2{\mbox e}^{iR\lambda/2\hbar} , \qquad 
C^\prime_2 = C_1 {\mbox e}^{iR\lambda/2\hbar} + C_2{\mbox e}^{-iR\lambda/2\hbar} . 
\end{equation}
In order to write the solution of these equations it is convenient to define
\begin{equation}
\label{def_varep}
\varepsilon = \frac{\lambda^{\alpha-1}}{2\pi\gamma} = \frac{1}{\alpha} \left(\frac{E}{U}\right)^{\frac{\alpha-1}{\alpha}} , 
\end{equation}
where $U$ was defined in Eq.(\ref{def_U}), in such a way that we have 
\begin{gather}
K_1(R) = \frac{2\pi\hbar \varepsilon}{W}[(\varepsilon \mu^{-1} +  \mathfrak{J}_\alpha(0))C^\prime_1 -  \mathfrak{J}_\alpha(\lambda R/\hbar) C^\prime_2] , \\
K_2(R) = \frac{2\pi \hbar \varepsilon}{\mu W} [(\varepsilon +  \mathfrak{J}_\alpha(0))C^\prime_2 -  \mathfrak{J}_\alpha(\lambda R/\hbar) C^\prime_1] , 
\end{gather}
where 
\begin{equation}
\label{eq.W}
W = (\varepsilon + \mathfrak{J}_\alpha(0))(\varepsilon \mu^{-1} +  \mathfrak{J}_\alpha(0)) -  (\mathfrak{J}_\alpha(\lambda R/\hbar))^2 . 
\end{equation}
Using $K_1(R)$ and $K_2(R)$ in Eq.(\ref{phi_dd_plus}) gives $\phi(p)$. Then, for $\psi(x)$, we have
\begin{equation}
\label{psi.double.delta.plus}
\begin{split}
\psi(x) & = C_1 {\mbox e}^{i\lambda x/\hbar} + C_2 {\mbox e}^{-i\lambda x/\hbar} \\
& \quad + \frac{1}{2\alpha W} [(\varepsilon \mu^{-1} +  \mathfrak{J}_\alpha(0))C^\prime_1 -  \mathfrak{J}_\alpha(\lambda R/\hbar) C^\prime_2]
 \Phi_\alpha \left(\frac{\lambda |x+R/2|}{\hbar}\right) \\
& \quad + \frac{1}{2\alpha W} [(\varepsilon +  \mathfrak{J}_\alpha(0))C^\prime_2 -  \mathfrak{J}_\alpha(\lambda R/\hbar) C^\prime_1]
\Phi_\alpha \left(\frac{\lambda |x-R/2|}{\hbar}\right) , 
\end{split}
\end{equation}
where we have expressed the result in terms of the function $\Phi_\alpha$ defined in Eq.(\ref{def_Phi}).

\section{Calculation of the Transmission Coefficients}

In order to calculate the transmission coefficients, we need to 
know the asymptotic bahaviour of the solutions. 
The asymptotic behaviour 
of Fox's $H$-function is given, if $\Delta > 0$, by Eq.(\ref{asymp1}) or Eq.(\ref{asymp2}) 
according to $\Delta^\ast > 0$ or $\Delta^\ast = 0$, respectively -- see 
Eq.(\ref{def_deltas}). In $\Phi_\alpha(\lambda |x|/\hbar)$ we have 
the difference between two Fox's $H$-functions of the form 
\begin{equation*}
H_{2,3}^{2,1} \left[ w^\alpha  \left| 
\begin{array}{l}
(1,1), (1,\mu) \\
(1,\alpha), (1,1), (1,\mu) \end{array} \right. \right] , 
\end{equation*}
for $\mu = (2+\alpha)/2$ and $\mu = (2-\alpha)/2$. 
In both cases we have $\Delta = \alpha > 0$, but 
$\Delta^\ast = 0$ for $\mu = (2+\alpha)/2$ and 
$\Delta^\ast > 0$ for $\mu = (2-\alpha)/2$. 
Therefore, using Eq.(\ref{asymp1}) when $\mu = (2-\alpha)/2$
and Eq.(\ref{asymp2}) when $\mu = (2+\alpha)/2$ we have, respectively,  
that 
\begin{gather}
 H_{2,3}^{2,1} \left[ w^\alpha  \left| 
\begin{array}{l}
(1,1), (1,(2+\alpha)/2) \\
(1,\alpha), (1,1), (1,(2+\alpha)/2) \end{array} \right. \right] = 
\frac{2 w}{\alpha} \sin{w} + \mathit{o}(1) , \quad |w| \rightarrow \infty , \\
H_{2,3}^{2,1} \left[ w^\alpha   \left| 
\begin{array}{l}
(1,1), (1,(2-\alpha)/2) \\
(1,\alpha), (1,1), (1,(2-\alpha)/2) \end{array} \right. \right] = 
\mathit{o}(1) , \qquad |w| \rightarrow \infty , 
\end{gather} 
and then 
\begin{equation}
\label{psi_asymp}
\Phi_\alpha\left(\frac{\lambda |x|}{\hbar}\right) = 
2\sin{\frac{\lambda |x|}{\hbar}} + \mathit{o}\big(|x|^{-1}\big) , \quad 
|x| \rightarrow \infty . 
\end{equation}

\subsection{Transmission Coefficient for the Delta Potential}

The behaviour of the wave function $\psi(x)$ given by Eq.(\ref{psi.delta.plus}) for $x\rightarrow \pm \infty$ is 
therefore 
\begin{equation}
\psi(x) = C_1 {\mbox e}^{i\lambda x/\hbar} + C_2 {\mbox e}^{-i\lambda x/\hbar} 
\pm \Omega_\alpha (C_1+C_2)\sin{\frac{\lambda x}{\hbar}} + \mathit{o}\big(x^{-1}\big) , \quad 
x\rightarrow \pm \infty . 
\end{equation}
or 
\begin{gather}
\label{psi_exp}
\psi(x) = A {\mbox e}^{i\lambda x/\hbar} + B {\mbox e}^{-i\lambda x/\hbar} + \mathit{o}\big(x^{-1}\big) , 
\quad x \rightarrow -\infty , \\
\psi(x) = C {\mbox e}^{i\lambda x/\hbar} + D {\mbox e}^{-i\lambda x/\hbar} + \mathit{o}\big(x^{-1}\big) , 
\quad x \rightarrow +\infty ,
\end{gather}
where we defined 
\begin{gather}
A = C_1 + i(C_1+C_2)\Omega_\alpha /2 , \quad B = C_2 - i(C_1+C_2)\Omega_\alpha /2  , \\
C = C_1 - i(C_1+C_2)\Omega_\alpha /2 , \quad D = C_2 + i(C_1+C_2)\Omega_\alpha /2  . 
\end{gather}
Now let us consider the situation of particles coming from the left and 
scattered by the delta potential. In this case $D = 0$ (no particles
coming from the right) and  
$B = r A$ and $C = t A$, where the reflexion $\mathcal{R}$ and transmission $\mathcal{T}$
coefficients are given by $\mathcal{R} = |r|^2$ and $\mathcal{T} = |t|^2$ (see, for 
example, \cite{Griffiths}). The result is 
\begin{equation}
\label{eq_r_t}
r = \frac{-i\Omega_\alpha}{1+i\Omega_\alpha} , \quad 
t = \frac{1}{1+i\Omega_\alpha} , 
\end{equation}
and then 
\begin{equation}
\label{ref_and_trans}
\mathcal{R} = \frac{\Omega_\alpha^2}{1+\Omega_\alpha^2} , \quad 
\mathcal{T} = \frac{1}{1+\Omega_\alpha^2} . 
\end{equation}
In Fig.(1) we show the behaviour of these coefficients for different 
values of $\alpha$. This plot and the other ones of this paper have 
been done by means of numerical integration of Eq.(\ref{jota.alpha}) 
using Mathematica 7.  

\begin{figure}[hbt]
\label{fig.2}
\begin{center}
\includegraphics[width=12cm]{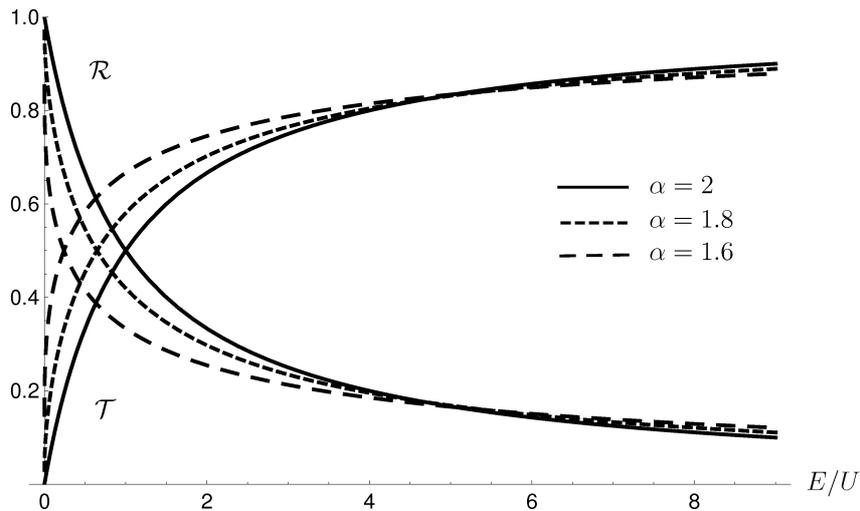}
\end{center}
\caption{Reflection and transmission coefficients as function of 
$E/U$, as given 
by Eq.(\ref{ref_and_trans}), for different values of $\alpha$.}
\end{figure}

We must note that transmission coefficient has a very interesting behaviour at zero energy. 
If we take the limit $E \rightarrow 0$ in the expression for $\Omega_\alpha$ in Eq.(\ref{def_omega}) 
we see that 
\begin{equation}
\lim_{E\rightarrow 0}\Omega_\alpha = -\tan\frac{\pi}{\alpha} , 
\end{equation}
and then for the transmission coefficient $\mathcal{T}$ we have 
\begin{equation}
\lim_{E\rightarrow 0} \mathcal{T} = \cos^2\frac{\pi}{\alpha} . 
\end{equation}
This is an unexpected and very interesting effect, 
which demands further interpretation (see Conclusions).

\subsection{Transmission Coefficient for the Double Delta Potential}

Let us introduce the following notations: 
\begin{equation}
\label{notation.double}
\mathcal{U} = \frac{\varepsilon \mu^{-1} + \mathfrak{J}_\alpha(0)}{\alpha W} , \qquad 
\mathcal{V} = \frac{\varepsilon + \mathfrak{J}_\alpha(0)}{\alpha W} , \qquad 
\mathcal{X} = \frac{\mathfrak{J}_\alpha(\lambda R /\hbar)}{\alpha W} . 
\end{equation}
The asymptotic behaviour of the wave function $\psi(x_)$ given by Eq.(\ref{psi.double.delta.plus}) 
for $x\rightarrow \pm \infty$ is 
therefore 
\begin{equation}
\begin{split}
\psi(x) & = C_1 {\mbox e}^{i\lambda x/\hbar} + C_2 {\mbox e}^{-i\lambda x/\hbar} 
+  (U C^\prime_1 -  X C^\prime_2) \sin{\frac{\lambda |x+R/2|}{\hbar}} \\
& \quad +
(V C^\prime_2 -  X C^\prime_1)
\sin{\frac{\lambda |x-R/2|}{\hbar}} +  \mathit{o}\big(x^{-1}\big) , \quad 
x\rightarrow \pm \infty , 
\end{split}
\end{equation}
or 
\begin{gather}
\label{psi_exp_2}
\psi(x) = A^\prime {\mbox e}^{i\lambda x/\hbar} + B^\prime {\mbox e}^{-i\lambda x/\hbar} + \mathit{o}\big(x^{-1}\big) , 
\quad x \rightarrow -\infty , \\
\psi(x) = C^\prime {\mbox e}^{i\lambda x/\hbar} + D^\prime {\mbox e}^{-i\lambda x/\hbar} + \mathit{o}\big(x^{-1}\big) , 
\quad x \rightarrow +\infty ,
\end{gather}
where we defined 
\begin{gather}
A^\prime = C_1 + M_1 , \qquad B^\prime = C_2 + M_2 , \\
C^\prime = C_1 - M_1 , \qquad D^\prime = C_2 - M_2 , 
\end{gather}
and 
\begin{gather}
M_1 = i(\rho C_1 + \sigma C_2 + i\tau C_2) , \\
M_2 = -i(\sigma C_1 + \rho C_2 - i\tau C_1) , 
\end{gather}
with 
\begin{equation}
\label{notation.aux}
\begin{split}
& \rho = \left(\frac{\mathcal{U}+\mathcal{V}}{2}\right) - \mathcal{X}\cos\frac{\lambda R}{\hbar} , \qquad 
\sigma = \left(\frac{\mathcal{U}+\mathcal{V}}{2}\right)\cos\frac{\lambda R}{\hbar} - \mathcal{X} , \\
& \phantom{\rho = \left(\frac{U+V}{2}\right) - X} \tau = \left(\frac{\mathcal{U}-\mathcal{V}}{2}\right)\sin\frac{\lambda R}{\hbar} . 
\end{split}
\end{equation}

As in the case of the delta potential, let us consider the situation of particles coming from the left and 
scattered by the double delta potential. In complete analogy we have $D^\prime = 0$ (no particles
coming from the right) and  
$B^\prime = r A^\prime$ and $C^\prime = t A^\prime$, where the reflexion $\mathcal{R}$ and transmission $\mathcal{T}$
coefficients are given by $\mathcal{R} = |r|^2$ and $\mathcal{T} = |t|^2$. The result is 
\begin{equation}
\label{eq_r_t}
r = \frac{2(\tau + i\sigma)}{(\rho^2 -\sigma^2-\tau^2-1) - 2 i \rho}  , \quad 
t = - \frac{(\rho^2-\sigma^2-\tau^2+1)}{(\rho^2 -\sigma^2-\tau^2-1) - 2 i \rho} . 
\end{equation}
and $\mathcal{R}$ and $\mathcal{T}$ can be written as 
\begin{equation}
\label{ref_and_trans_double}
\mathcal{R} = \frac{4(\sigma^2+\tau^2)}{(\rho^2-\sigma^2-\tau^2+1)^2+ 4(\sigma^2 + \tau^2)} , \quad 
\mathcal{T} = \frac{(\rho^2-\sigma^2-\tau^2+1)^2}{(\rho^2-\sigma^2-\tau^2+1)^2+ 4(\sigma^2 + \tau^2)} . 
\end{equation}
We can simplify these expressions a little bit once we note that 
\begin{equation}
\rho^2 -\sigma^2 -\tau^2 = \frac{\sin^2{(\lambda R/\hbar)}}{\alpha^2 W} , 
\end{equation}
and then 
\begin{equation}
\label{final_coeff}
\mathcal{R} = \frac{\Delta_\alpha^2}{1+\Delta_\alpha^2} , \qquad \mathcal{T} = \frac{1}{1+\Delta_\alpha^2} , 
\end{equation}
with 
\begin{equation}
\label{def_delta_alpha}
\Delta_\alpha^2 = \frac{4\alpha^4 W^2(\sigma^2+\tau^2)}{(\alpha^2 W + \sin^2{(\lambda R/\hbar)})^2} . 
\end{equation}

In Fig.(2) we show the behaviour of the transmission coefficient for different 
values of $\alpha$ and in Fig.(3) for different values of $\mu$. 

\begin{figure}[hbt]
\label{fig.3}
\begin{center}
\includegraphics[width=12cm]{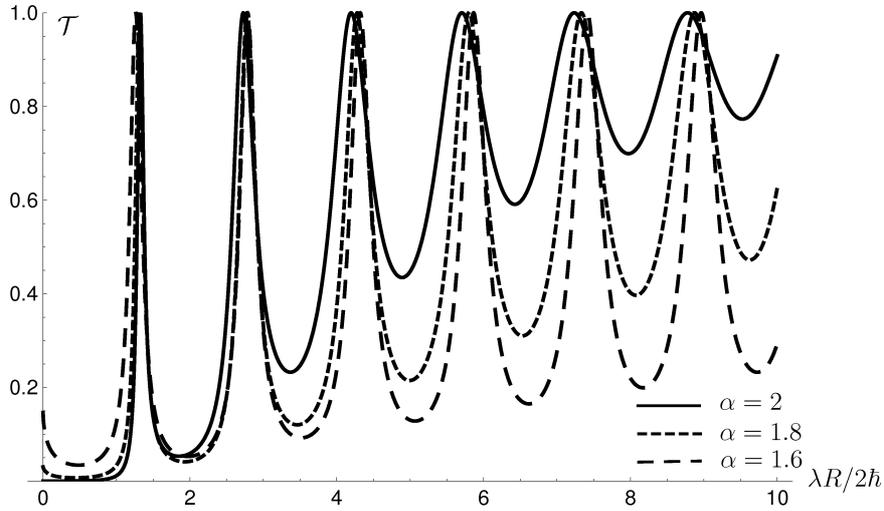}
\end{center}
\caption{Transmission coefficients as function of 
$\lambda R/2\hbar$, as given 
by Eq.(\ref{final_coeff}), for different values of $\alpha$, 
when $2\pi \gamma=10$ and $\mu = 1$.}
\end{figure}

\begin{figure}[hbt]
\label{fig.4}
\begin{center}
\includegraphics[width=15cm]{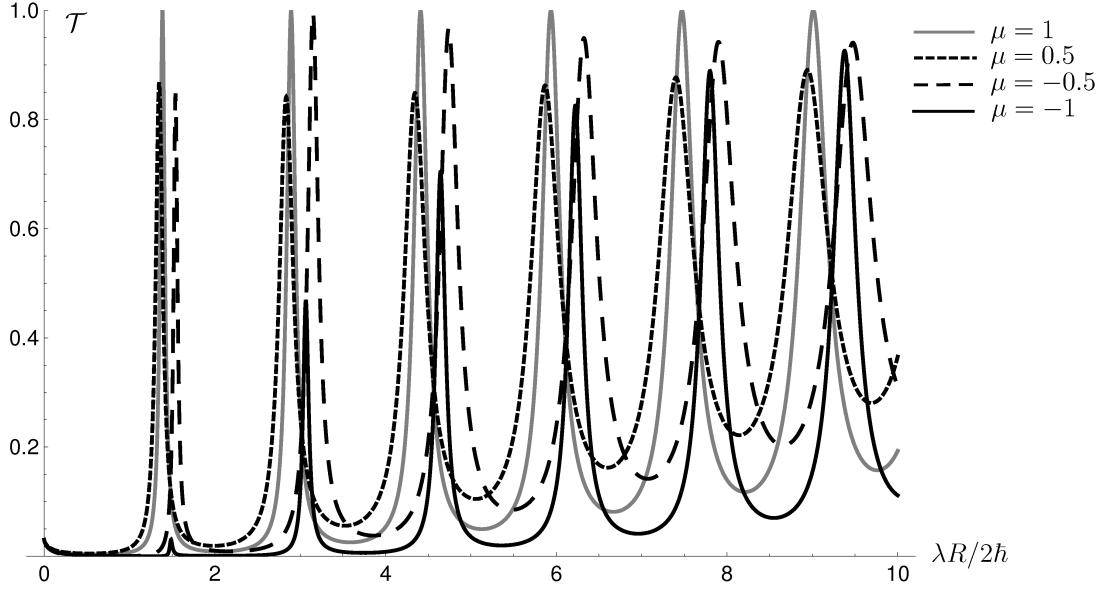}
\end{center}
\caption{Transmission coefficients as function of 
$\lambda R/2\hbar$, as given 
by Eq.(\ref{final_coeff}), for different values of $\mu$, 
when $2\pi \gamma=20$ and $\alpha = 1.8$.}
\end{figure}

As in the case of the delta potential, we have a very interesting behaviour 
for these coefficients when $E\rightarrow 0$. Firstly, using Eq.(\ref{JExpansion}) 
in the Appendix B, we have 
\begin{equation}
\mathfrak{J}_\alpha(\lambda R/\hbar) = \mathfrak{J}_\alpha(0) + A_1 \lambda^{\alpha-1} + A_2 \lambda^2 + \mathcal{O}(\lambda^{3\alpha-1}) , 
\end{equation}
where 
\begin{equation}
A_1 = \frac{(R/\hbar)^{\alpha-1}}{2\Gamma(\alpha)\cos{(\pi\alpha/2)}} , \qquad 
A_2 = \frac{(R/\hbar)^2\cot{(3\pi/\alpha)}}{2\alpha} . 
\end{equation}
Using this, the expression Eq.(\ref{eq.W}) for $W$ gives 
\begin{equation}
W = \mathfrak{J}_\alpha(0)(H (1+\mu^{-1})-2 A_1) \lambda^{\alpha-1} + (H^2 \mu^{-1} - A_1^2) 
\lambda^{2(\alpha-1)} - 2\mathfrak{J}_\alpha(0) A_2 \lambda^2 + \mathcal{O}(\lambda^{\alpha+1}) , 
\end{equation}
where $H = 1/2\pi \gamma$. Then, with some calculations, we obtain that 
\begin{equation}
4\alpha^2 W^2(\sigma^2+\tau^2) = B_1 \lambda^{2(\alpha-1)} + B_2 \lambda^{\alpha+1} + \mathcal{O}(\lambda^{2\alpha}) , 
\end{equation}
with 
\begin{equation}
B_1 = (H(1+\mu^{-1})-2A_1)^2 , \qquad B_2 = -2(H(1+\mu^{-1})-2A_1)(2A_2+\mathfrak{J}_\alpha(0)(R/\hbar)^2) , 
\end{equation}
and 
\begin{equation}
\alpha^2 W^2(\rho^2-\sigma^2-\tau^2+1) = 
B_1^\prime \lambda^{2(\alpha-1)} + B_2^\prime \lambda^{4(\alpha-1)} + 
B_3^\prime \lambda^{\alpha+1} + \mathcal{O}(\lambda^{2\alpha}) , 
\end{equation}
with 
\begin{equation}
\begin{split}
& B_1^\prime = \alpha^2 \mathfrak{J}^2_\alpha(0) (H(1+\mu^{-1})-2A_1)^2 , \qquad 
B_2^\prime = \alpha^2(H^2 \mu^{-1}-A_1)^2 , \\
& B_3^\prime = -2 \alpha \mathfrak{J}_\alpha(0)(H(1+\mu^{-1})-2A_1)(2\alpha^2 \mathfrak{J}_\alpha(0) A_2 -(R/\hbar)^2) . 
\end{split}
\end{equation}
Using these results we can easily see that, for $E/D_\alpha = \lambda^\alpha \rightarrow 0$, 
\begin{equation}
\lim_{E\rightarrow 0} \Delta_\alpha^2 = \frac{1}{\alpha^2 \mathfrak{J}^2_\alpha(0)} , 
\end{equation}
and, since $\mathfrak{J}_\alpha(0) = -(1/\alpha)\cot{(\pi/\alpha)}$, that   
\begin{equation}
\lim_{E\rightarrow 0} T = \cos^2\frac{\pi}{\alpha} . 
\end{equation}
This the same result we obtained for the zero energy limit of the transmission 
coefficient for a single delta potential. Moreover, this limit does not depends 
on $\mu$, which is the parameter that relates the two delta functions in the
potential in Eq.(\ref{delta.potential}). 
Again, this is a very interesting 
and unexpected result.

\section{Conclusions}

The tunneling effect in fractional quantum mechanics 
has some very interesting properties which are not 
observed in the usual $\alpha = 2$ quantum mechanics. 
The most interesting is the presence of tunneling 
through delta and double delta potentials  
even at zero energy. Moreover, in the case 
of the double delta potential, this zero energy 
tunneling is independent of the relation of the
two delta functions. Let us give a possible 
explanation to these results. 

FQM was defined from the point of view of path integrals. 
As well-known, in this approach, 
when the sum is taken over paths of Brownian motion type, 
we have the standard quantum mechanics. On the other hand, 
in FQM the sum is taken over the paths of L\'evy flights \cite{Laskin1}, 
which are generalizations of Brownian motion, and such that the 
corresponding probability distribution has infinite variance. 
By means of L\'evy flights, there is a not negligible probability 
of a particle reaching faraway points in 
a single jump, in contrast to a random
walk of Brownian motion type \cite{Dubkov}.  

In FQM, an uncertainty principle still holds, but
with an appropriated modification, that is, we have \cite{Laskin2} 
\begin{equation}
\label{UP}
\langle |\Delta x|^\mu\rangle^{1/\mu} \langle |\Delta p|^\mu \rangle^{1/\mu} 
> \frac{\hbar}{(2\alpha)^{1/\mu}} , \quad \mu < \alpha , \quad 1 < \alpha \leq 2 , 
\end{equation}
and such that in the standard quantum mechanics limit we can also have $\mu = \alpha = 2$. 
Thus, even when $E = 0$, the particle can 
have energy $\Delta E = \langle |\Delta p|^\mu \rangle^{2/\mu}/2m$ 
and momentum $\langle |\Delta p|^\mu \rangle^{1/\mu}$. 
This may not be enough for tunneling through a delta potential 
in the standard case, but in the fractional one, where long jumps 
of L\'evy flights enter the sum in the path integral, this may 
be responsible for the tunneling with probability  
$\cos^2{(\pi/\alpha)}$. 

\bigskip

\noindent \textbf{\sffamily Acknowledgements:} We are grateful to the 
referees for many valuable suggestions.

\appendix 

\section{Fox's $H-$Function} 
The Fox's $H-$function, also known as $H-$function or Fox's function, was introduced in the literature as an 
integral of Mellin-Barnes type \cite{Mathai}.

Let $m$, $n$, $p$ and $q$ be integer numbers. Consider the function
\begin{equation}
\label{eq.fox.1}
\Lambda (s) = \displaystyle \frac{\displaystyle \prod_{i=1}^m \Gamma(b_i + B_i \,s)  \prod_{i=1}^n \Gamma(1- a_i - A_i \,s)
}{\displaystyle \prod_{i=m+1}^q \Gamma(1 - b_i - B_i \,s)  \prod_{i=n+1}^p \Gamma(a_i + A_i \,s)} 
\end{equation}
with $1 \leq m \leq q$ and $0 \leq n \leq p$. The coefficients $A_i$ and $B_i$ are positive real numbers; $a_i$ and 
$b_i$ are complex parameters.

The Fox's $H-$function, denoted by, 
\begin{equation}
H_{p,q}^{m,n}(x) = 
H_{p,q}^{m,n} \left( x \left| 
\begin{array}{c}
(a_p,A_p) \\
(b_q,B_q)
\end{array} \right. \right) =
H_{p,q}^{m,n} \left[ x \left| 
\begin{array}{c}
(a_1,A_1), \cdots, (a_p,A_p) \\
(b_1,B_1), \cdots, (b_q,B_q)
\end{array} \right. \right]
\end{equation}
is defined as the inverse Mellin transform, i.e., 
\begin{equation}
\label{eq.fox.2}
H_{p,q}^{m,n} (x) = \frac{1}{2\pi i} \int_{L} \Lambda (s) \, x^{-s} \, {\mbox{d}}s
\end{equation}
where $\Lambda (s)$ is given by Eq.(\ref{eq.fox.1}), and the contour $L$ runs from $L-i\infty$ to $L+i\infty$ separating 
the poles of $\Gamma(1-a_i-A_i \, s)$, $(i=1,\ldots ,n)$ from those of $\Gamma(b_i + B_i \, s)$, $(i=1,\ldots ,m)$. 
The complex parameters $a_i$ and $b_i$ are taken with the imposition that no poles in the integrand coincide.

There are some interesting properties associated with the Fox's $H-$function. 
We consider here the following ones: 

\subsubsection*{P.1. Change the independent variable}
Let $c$ be a positive constant. We have
\begin{equation}
\label{Prop1}
H_{p,q}^{m,n} \left[ x \left| 
\begin{array}{c}
(a_p,A_p) \\
(b_q,B_q)
\end{array} \right. \right] = c \, 
H_{p,q}^{m,n} \left[ x^c \left| 
\begin{array}{c}
(a_p,c\,A_p) \\
(b_q,c\,B_q)
\end{array} \right. \right]. 
\end{equation}
To show this expression one introduce a change of variable $s \to c\,s$ in the integral of inverse Mellin transform.

\subsubsection*{P.2. Change the first argument}
Set $\alpha \in \mathbb{R}$. Then we can write
\begin{equation}
\label{Prop2}
x^{\alpha} H_{p,q}^{m,n} \left[ x \left| 
\begin{array}{c}
(a_p,A_p) \\
(b_q,B_q)
\end{array} \right. \right] =  
H_{p,q}^{m,n} \left[ x \left| 
\begin{array}{c}
(a_p+ \alpha A_p,A_p) \\
(b_q+ \alpha B_q,B_q)
\end{array} \right. \right]. 
\end{equation}
To show this expression first we introduce the change $a_p \to a_p + \alpha A_p$ and take $s \to s - \alpha$ in the 
integral of inverse Mellin transform.

\subsubsection*{P.3. Lowering of Order}
If the first factor $(a_1,A_1)$ is equal to the last one, $(b_q,B_q)$, we have
\begin{equation}  
H_{p,q}^{m,n} \left[ x \left| 
\begin{array}{l}
(a_1,A_1), \cdots, (a_p,A_p) \\
(b_1,B_1),\cdots, (b_{q - 1},B_{q - 1}) (a_1,A_1)
\end{array} \right. \right] = 
H_{p-1,q-1}^{m,n-1} \left[ x \left| 
\begin{array}{l}
(a_2,A_2), \cdots, (a_p,A_p) \\
(b_1,B_1), \cdots, (b_{q-1},B_{q-1})
\end{array} \right. \right]. \label{Ap3}
\end{equation}
To show this identity is sufficient to simplify the common arguments in the Mellin-Barnes integral.

\subsubsection*{P.4. Asymptotic Expansions}
The asymptotic expansions for Fox's $H$-functions have been studied in \cite{Braaskma}. 
Let $\Delta$ and $\Delta^\ast$ be defined as 
\begin{equation}
\label{def_deltas}
\Delta = \sum_{i=1}^q B_i - \sum_{i=1}^p A_i , \qquad 
\Delta^\ast = \sum_{i=1}^n A_i - \sum_{i=n+1}^p A_i + \sum_{i=1}^m B_i -\sum_{i=m+1}^q B_i . 
\end{equation}
If $\Delta > 0$ and $\Delta^\ast > 0$ we have \cite{Kilbas}
\begin{equation}
\label{asymp1}
H^{m,n}_{p,q}(x) = \sum_{r=1}^n \big[ h_r x^{(a_r-1)/A_r} + \mathit{o}\big( x^{(a_r-1)/A_r}\big) \big] , \qquad |x|\rightarrow \infty . 
\end{equation}
where 
\begin{equation}
h_r = \frac{1}{A_r}\frac{\prod_{j=1}^m \Gamma(b_j+(1-a_r)B_j/A_r) \prod_{j=1,j\neq r}^n \Gamma(1-a_j-(1-a_r)A_j/A_r)}
{\prod_{j=n+1}^p \Gamma(a_j -(1-a_r)A_j/A_r) \prod_{j=m+1}^q \Gamma(1-b_j-(1-a_r)B_j/A_r)} , 
\end{equation}
and if $\Delta > 0$ and $\Delta^\ast = 0$ we have \cite{Kilbas}
\begin{equation}
\label{asymp2}
\begin{split}
H^{m,n}_{p,q}(x) & = \sum_{r=1}^n \big[ h_r x^{(a_r-1)/A_r} + \mathit{o}\big( x^{(a_r-1)/A_r}\big) \big] \\
& \quad + A x^{(\nu+1/2)/\Delta} \big( c_0 \exp[i(B + C x^{1/\Delta})] - d_0 \exp[-i(B+C x^{1/\Delta})]\big) \\
& \quad + \mathit{o}\big(x^{(\nu+1/2)/|\Delta|}\big) , \qquad |x| \rightarrow \infty , 
\end{split}
\end{equation}
where 
\begin{gather*}
c_0 = (2\pi i)^{m+n-p}\exp\big[\pi i \big(\sum_{r=n+1}^p a_r - \sum_{j=1}^m b_j\big)\big] , \\
d_0 = (-2\pi i)^{m+n-p}\exp\big[-\pi i \big(\sum_{r=n+1}^p a_r - \sum_{j=1}^m b_j\big)\pi i\big] , \\
A = \frac{1}{2\pi i \Delta}(2\pi)^{(p-q+1)/2} \Delta^{-\nu} \prod_{r=1}^p A_r^{-a_r+1/2} \prod_{j=1}^q B_j^{b_j-1/2} 
\bigg(\frac{\Delta^\Delta}{\delta}\bigg)^{(\nu+1/2)/\Delta} , 
\end{gather*}
\begin{gather*}
B = \frac{(2\nu+1)\pi}{4} , \qquad C = \bigg(\frac{\Delta^\Delta}{\delta}\bigg)^{1/\Delta} , \\
\delta = \prod_{l=1}^p |A_l|^{-A_l} \prod_{j=1}^q |B_j|^{B_j} , \qquad 
\nu = \sum_{j=1}^q b_j - \sum_{j=1}^p a_j + \frac{p-q}{2} . 
\end{gather*}

\subsubsection*{P.5. Series Expansion}

In \cite{Mathai} we can see that 
in some cases there is a series expansion for Fox's $H$-function. 
For example, when the poles of $\prod_{j=1}^m \Gamma(b_j + B_j s)$ are simple, 
we can write 
\begin{equation}
\label{fox_series}
H^{m,n}_{p,q}(x) = \sum_{j=1}^m \sum_{\nu=0}^{\infty} h_{j\nu} x^{(b_j+\nu)/B_j} , 
\end{equation}
where 
\begin{equation}
\label{fox_series_coeff}
h_{j\nu} = \frac{(-1)^\nu}{\nu! B_j}\frac{ \prod_{i=1,i\neq j}^m \Gamma(b_i-B_i(b_j+\nu)/B_j) 
\prod_{i=1}^n \Gamma(1-a_i + A_i(b_j+\nu)/B_j)}{\prod_{i=m+1}^q \Gamma(1-b_i+B_i(b_j+\nu)/B_j) 
\prod_{i=n+1}^p \Gamma(a_i-A_i(b_j+\nu)/B_j)} . 
\end{equation}

\section{Calculation of the Integral in Eq.(\ref{jota.alpha})}

Let $\mathfrak{J}_\alpha(w)$ be given by 
\begin{equation}
\mathfrak{J}_\alpha(w) = \frac{1}{\pi} \int_0^{+\infty}\frac{\cos{w y}}{y^\alpha-1}{\mbox d}y . 
\end{equation}
Taking the Mellin transform we have that 
\begin{equation}
\mathcal{M}_w[\mathfrak{J}_\alpha(w)](z) = \frac{1}{\pi} \Gamma(z) \cos{\frac{\pi z}{2}} \int_0^{+\infty} \frac{y^{-z}}{y^\alpha-1}{\mbox d}y . 
\end{equation}
This last integral is given by formula 3.241.3 (pg.322) of \cite{Grad}, that is, 
\begin{equation}
\label{grad3.241.3}
\int_0^{+\infty} \frac{x^{\mu-1}}{1-x^\nu}{\mbox d}x = \frac{\pi}{\nu}\cot{\frac{\mu \pi}{\nu}} ,  
\end{equation}
where the integration is understood as the Cauchy principal value\footnote{We remember that in the inversion of
the Fourier transform the integration is to be done in the sense of the Cauchy principal value \cite{Churchill}.}. 
Therefore we have 
\begin{equation}
\mathcal{M}_w[\mathfrak{J}_\alpha(w)](z) = -\frac{1}{\alpha}\Gamma(z) \sin{\frac{\pi(1-z)}{2}} \cot{\frac{\pi(1-z)}{\alpha}} . 
\end{equation}
Using the relation $2\sin{A}\cos{B} = \sin{(A+B)}+\sin{(A-B)}$ and writing the sine function in terms of the product of
gamma functions we can write that 
\begin{equation*}
\begin{split}
\mathcal{M}_w[\mathfrak{J}_\alpha(w)](z) & = -\frac{1}{2\alpha}\frac{\Gamma(z)\Gamma\big(\frac{1-z}{\alpha}\big)\Gamma\big(1-\frac{1-z}{\alpha}\big)} 
{\Gamma\big((1-z)\frac{(2+\alpha)}{2\alpha}\big)\Gamma\big(1-(1-z)\frac{(2+\alpha)}{2\alpha}\big)} \\
& \quad +\frac{1}{2\alpha}\frac{\Gamma(z)\Gamma\big(\frac{1-z}{\alpha}\big)\Gamma\big(1-\frac{1-z}{\alpha}\big)} 
{\Gamma\big((1-z)\frac{(2-\alpha)}{2\alpha}\big)\Gamma\big(1-(1-z)\frac{(2-\alpha)}{2\alpha}\big)} = F_2(z) . 
\end{split}
\end{equation*}
Taking the inverse Mellin transform and using the definition of the Fox's $H$-function we have that
\begin{equation*}
\begin{split}
\mathfrak{J}_\alpha(w) & = -\frac{1}{2\alpha} 
H_{2,3}^{2,1} \left[ w \left| 
\begin{array}{l}
(1-1/\alpha,1/\alpha), (1-(2+\alpha)/2\alpha,(2+\alpha)/2\alpha)   \\
(0,1), (1-1/\alpha,1/\alpha), (1-(2+\alpha)/2\alpha,(2+\alpha)/2\alpha) \end{array} \right. \right] + \\
& \quad + \frac{1}{2\alpha} 
H_{2,3}^{2,1} \left[ w \left| 
\begin{array}{l}
(1-1/\alpha,1/\alpha), (1-(2-\alpha)/2\alpha,(2-\alpha)/2\alpha)   \\
(0,1), (1-1/\alpha,1/\alpha), (1-(2-\alpha)/2\alpha,(2-\alpha)/2\alpha) \end{array} \right. \right] . 
\end{split}
\end{equation*}
Using the properties given by Eqs.(\ref{Prop1},\ref{Prop2}) and replacing $w$ by 
$|w|$ since $\mathfrak{J}_\alpha(-w) = \mathfrak{J}_\alpha(w)$ we obtain 
\begin{equation}
\label{JFox}
\begin{split}
\mathfrak{J}_\alpha(w) & = -\frac{1}{2|w|} 
H_{2,3}^{2,1} \left[ |w|^\alpha \left| 
\begin{array}{l}
(1,1), (1,(2+\alpha)/2)   \\
(1,\alpha), (1,1), (1,(2+\alpha)/\alpha) \end{array} \right. \right] + \\
& \quad + \frac{1}{2|w|} 
H_{2,3}^{2,1} \left[ |w|^\alpha \left| 
\begin{array}{l}
(1,1), (1,(2-\alpha)/2)   \\
(1,\alpha), (1,1), (1,(2-\alpha)/2) \end{array} \right. \right] . 
\end{split}
\end{equation}
 
Let us see what happens in the particular case $\alpha = 2$. From the definition of Fox's $H$-function we 
can see that 
\begin{equation}
H_{2,3}^{2,1} \left[ |w|^2 \left| 
\begin{array}{l}
(1,1), (1,0)   \\
(1,2), (1,1), (1,0) \end{array} \right. \right] = 0 
\end{equation}
and that 
\begin{equation*}
H_{2,3}^{2,1} \left[ w^2 \left| 
\begin{array}{l}
(1,1), (1,2)   \\
(1,2), (1,1), (1,2) \end{array} \right. \right] = 
H_{1,2}^{1,1} \left[ w^2 \left| 
\begin{array}{l}
(1,1)   \\
(1,1), (1,2) \end{array} \right. \right] = 
w^2 H_{1,2}^{1,1} \left[ w^2 \left| 
\begin{array}{l}
(0,1)   \\
(0,1), (-1,2) \end{array} \right. \right] . 
\end{equation*}
But \cite{Mathai}
\begin{equation}
H_{1,2}^{1,1} \left[ -z \left| 
\begin{array}{l}
(0,1)   \\
(0,1), (1-b,a) \end{array} \right. \right] = E_{a,b}(z) , 
\end{equation}
where $E_{a,b}(z)$ is the two-parameter Mittag-Leffler function. 
However, it is known \cite{Podlubny} that 
\begin{equation}
E_{2,2}(z) = \frac{\sinh{\sqrt{z}}}{\sqrt{z}} . 
\end{equation}
Consequently, we have 
\begin{equation}
H_{2,3}^{2,1} \left[ w^2 \left| 
\begin{array}{l}
(1,1), (1,2)   \\
(1,2), (1,1), (1,2) \end{array} \right. \right] = |w|^2 E_{2,2}(-|w|^2) = |w|\sin{|w|} . 
\end{equation}
Then for $\alpha = 2$ we have  
\begin{equation}
\label{jota.two}
\mathfrak{J}_2\left(\frac{\lambda x}{\hbar}\right) = -\frac{1}{2}\sin\frac{\lambda |x|}{\hbar} , 
\end{equation}
and 
\begin{equation}
\label{eq.fourier.inv.2.1}
\int_{-\infty}^{+\infty}\frac{{\mbox e}^{ipx/\hbar}}{|p|^2-\lambda^2} {\mbox d}p  = 
-\frac{\pi}{\lambda}\sin{\frac{\lambda |x|}{\hbar}} .
\end{equation}

We are also interested in the expression of $\mathfrak{J}_\alpha(w)$ for small $w$. 
From Eq.(\ref{JFox}) we see that we need to know the behaviour of 
$$
H_{2,3}^{2,1} \left[ |w|^\alpha \left| 
\begin{array}{l}
(1,1), (1,\mu)   \\
(1,\alpha), (1,1), (1,\mu) \end{array} \right. \right]
$$
for small $w$. This is given by the series expansion 
from Eq.(\ref{fox_series}), which gives 
\begin{equation}
\begin{split}
& H_{2,3}^{2,1} \left[ z \left| 
\begin{array}{l}
(1,1), (1,\mu)   \\
(1,\alpha), (1,1), (1,\mu) \end{array} \right. \right] \\
& \quad = \frac{\Gamma(1-1/\alpha)\Gamma(1/\alpha)}{\Gamma(1-\mu/\alpha)\Gamma(\mu/\alpha)}\frac{z^{1/\alpha}}{\alpha} 
-\frac{\Gamma(1-2/\alpha)\Gamma(2/\alpha)}{\Gamma(1-2\mu/\alpha)\Gamma(2\mu/\alpha)}\frac{z^{2/\alpha}}{\alpha} \\
& \qquad + 
\frac{\Gamma(1-3/\alpha)\Gamma(3/\alpha)}{\Gamma(1-3\mu/\alpha)\Gamma(3\mu/\alpha)}\frac{z^{3/\alpha}}{2\alpha} + 
\mathcal{O}(z^{4/\alpha}) \\
& \qquad + \frac{\Gamma(1-\alpha)\Gamma(1)}{\Gamma(1-\mu)\Gamma(\mu)}z - 
\frac{\Gamma(1-2\alpha)\Gamma(2)}{\Gamma(1-2\mu)\Gamma(2\mu)}z^2 + 
\frac{\Gamma(1-3\alpha)\Gamma(3)}{\Gamma(1-3\mu)\Gamma(3\mu)}\frac{z^3}{2} + 
\mathcal{O}(z^4) . 
\end{split}
\end{equation}
Using this in Eq.(\ref{JFox}) we arrive, after some manipulations, 
to 
\begin{equation}
\label{JExpansion}
\mathfrak{J}_\alpha(w) = \mathfrak{J}_\alpha(0) + \frac{1}{2\Gamma(\alpha)\cos{(\pi \alpha/2)}}w^{\alpha-1} + 
\frac{\cot{(3\pi/\alpha)}}{2\alpha} w^2 + \mathcal{O}(w^{3\alpha-1}) , 
\end{equation}
where 
\begin{equation}
\label{EqJ0} 
\mathfrak{J}_\alpha(0) = -\frac{1}{\alpha} \cot\frac{\pi}{\alpha} . 
\end{equation}

\section{The Limit $\alpha = 2$}

Let us calculate the transmission coefficients in 
the standard quantum mechanical limit and see 
that we recover the usual results. 

Firstly, let us consider
the delta potential. The transmission coefficient is given by 
Eq.(\ref{ref_and_trans}), so that we need to calculate $\Omega_2$ 
in this case, where $\Omega_\alpha$ is given by Eq.(\ref{def_omega}). 
The result is that 
\begin{equation}
\Omega_2 = \left(\frac{E}{U}\right)^{-1} , \qquad U = \frac{m V_0^2}{2\hbar^2} , 
\end{equation}
where we used the definition of $U$ in Eq.(\ref{def_U}) and $D_2 = 1/2m$, 
and such that 
\begin{equation}
\mathcal{T} = \frac{1}{1+ (mV_0^2/2\hbar^2 E)} , 
\end{equation}
which is the well-known result \cite{Griffiths}. 

Now let us consider the transmission coefficient for the double 
delta potential. Let us also consider the case $\mu = 1$ since
the result in this case is well-known \cite{Galindo}. In order
to calculate $\Delta_2$ in Eq.(\ref{final_coeff}) we need Eq.(\ref{jota.two}), 
which by the way gives 
\begin{equation}
\mathfrak{J}_2(0) = 0 . 
\end{equation}
When $\mu = 1$ we have 
\begin{equation}
\mathcal{U} = \mathcal{V} = \frac{1}{2W} , \qquad \tau = 0 , 
\end{equation}
and then 
\begin{equation}
\Delta_2^2 = \frac{64 W^2 \sigma^2}{[4W + \sin^2(\lambda R/\hbar)]^2} . 
\end{equation}
But in this case Eq.(\ref{eq.W}) together with Eq.(\ref{jota.two}) gives 
\begin{equation}
4W + \sin^2(\lambda R/\hbar) = 4\epsilon^2 . 
\end{equation}
From Eq.(\ref{notation.double}) and Eq.(\ref{notation.aux}) we also have that 
\begin{equation}
4W^2 \sigma^2 = \left( \epsilon \cos{\frac{\lambda R}{\hbar}} 
+ \frac{1}{2}\sin\frac{\lambda R}{\hbar} \right)^2 . 
\end{equation}
Then for $\Delta_2$ we have
\begin{equation}
\Delta_2^2 = \left(\frac{1}{\epsilon}\cos\frac{\lambda R}{\hbar} + 
\frac{1}{2\epsilon^2} \sin\frac{\lambda R}{\hbar}\right)^2 . 
\end{equation}
Let us change the notation a little bit in order to compare 
with the standard result in the literature. Let us define 
\begin{equation}
\mu_0  = \frac{\lambda R}{2\hbar} , \qquad \beta = \frac{\mu_0 }{\epsilon} .
\end{equation}
Using this notation in the above expression for $\Delta_2$, 
we have for $\mathcal{T}$ 
in Eq.(\ref{final_coeff}) that 
\begin{equation}
\mathcal{T} = \frac{\mu_0^4}{\mu_0^4 + [\beta \mu_0 \cos{2\mu_0} 
+ (\beta^2/2)\sin{2\mu_0}]^2} , 
\end{equation}
which is the result in \cite{Galindo}, pg 160 (where $\mu_0 = \mu$).

\section{The Boundary Conditions}

In \cite{EFJ} we have discussed how the Riesz fractional derivative 
can be written in terms of the Riesz potentials, that is, 
for $0 < \alpha < 1$ we have 
\begin{equation}
\label{factorriesz}
(-\triangle)^{\alpha/2}\psi(x) = 
\frac{{\mbox d}\hspace{1ex}}{{\mbox d}x}\tilde{\mathcal{R}}^{1-\alpha} \psi(x) , 
\end{equation}
and for $1 < \alpha < 2$ we have 
\begin{equation}
\label{factorriesz2}
(-\triangle)^{\alpha/2}\psi(x)  = -\frac{{\rm d^2}\hspace{1ex}}{{\mbox d}x^2}{\mathcal{R}}^{2-\alpha} \psi(x) , 
\end{equation}
where  $\mathcal{R}^{\alpha^\prime} \psi(x)$ 
is the Riesz potential of $\psi(x)$ of order ${\alpha^\prime}$ given by \cite{Butzer}
\begin{equation}
\label{rieszpot}
\mathcal{R}^{\alpha^\prime} \psi(x) = \frac{1}{2\Gamma({\alpha^\prime})
\cos{({\alpha^\prime}\pi/2)}}\int_{-\infty}^{+\infty} 
\frac{\psi(\xi)}{|x-\xi|^{1-{\alpha^\prime}}}{\mbox d}\xi , 
\end{equation}
for $0 < {\alpha^\prime} <1$, and $\tilde{\mathcal{R}}^{\alpha^\prime}\psi(x)$ its conjugated 
Riesz potential given by 
\begin{equation}
\label{conjrieszpot}
\tilde{\mathcal{R}}^{\alpha^\prime} \psi(x) = \frac{1}{2\Gamma({\alpha^\prime})
\sin{({\alpha^\prime}\pi/2)}}\int_{-\infty}^{+\infty} 
\frac{{\operatorname{sign}(x-\xi)}\psi(\xi)}{|x-\xi|^{1-{\alpha^\prime}}}{\mbox d}\xi . 
\end{equation}

If we use these expressions for the Riesz fractional derivative in the FSE for
the delta potential $V(x) = V_0 \delta(x)$ and integrate as usual from $-\epsilon$ 
to $+\epsilon$ and take the limit $\epsilon \rightarrow 0$ we obtain that 
\begin{equation}
\label{bc1}
\frac{{\mbox d}\hspace{1ex}}{{\mbox d}x} \mathcal{R}^{2-\alpha}\psi(x)\bigg|_{0^+} - 
\frac{{\mbox d}\hspace{1ex}}{{\mbox d}x} \mathcal{R}^{2-\alpha}\psi(x)\bigg|_{0^-} = 
\frac{V_0}{\hbar^\alpha D_\alpha}\psi(0) . 
\end{equation}
This condition and the continuity one $\psi(0^-) = \psi(0^+)$ are 
the boundary conditions to be satisfied by $\psi(x)$. 

There is an 
important point to be noted here: the expression 
$$
\frac{{\mbox d}\hspace{1ex}}{{\mbox d}x} \mathcal{R}^{2-\alpha}\psi(x) = 
\frac{{\mbox d}\hspace{1ex}}{{\mbox d}x} \mathcal{R}^{1-(\alpha-1)}\psi(x) 
$$
is \textit{not} the Riesz fractional derivative of order $\alpha-1$, which 
is given, for $0 < \alpha - 1 < 1$, by 
$$
\frac{{\mbox d}\hspace{1ex}}{{\mbox d}x} \tilde{\mathcal{R}}^{1-(\alpha-1)}\psi(x) . 
$$
Therefore, it is wrong to write the condition (\ref{bc1}) as 
\begin{equation}
\label{bc1.wrong} 
(-\triangle)^{(\alpha-1)/2}\psi(x)\big|_{0^+} - 
(-\triangle)^{(\alpha-1)/2}\psi(x)\big|_{0^-} =
\frac{V_0}{\hbar^\alpha D_\alpha}\psi(0) , 
\end{equation}
with $(-\triangle)^{\alpha^\prime/2}$ being the Riesz 
fractional derivative. Maybe in another version of FQM 
involving a fractional derivative defined in another 
sense, it can holds one such condition, 
but this is not the case when it comes to the Riesz fractional 
derivative. 
In \cite{Lin} the authors have used a boundary condition of the 
above type in their attempt to solve the problem for the double delta 
potential. Besides the already discussed problem with their local approach, it also 
seems that the use of 
that boundary condition is not justified. 

Let us consider the solution in the case of the delta potential 
and show that it satisfy Eq.(\ref{bc1}). In \cite{EFJ} we have 
seen that 
\begin{equation}
\frac{{\mbox d}\hspace{1ex}}{{\mbox d}x}\mathcal{R}^{2-\alpha}\psi(x) = 
\frac{i}{2\pi \hbar^\alpha}\int_{-\infty}^\infty {\rm e}^{-ipx/\hbar} |p|^{\alpha-1} 
\operatorname{sign}(p) \phi(p) \, {\mbox d}p . 
\end{equation}
Using $\phi(p)$ given by Eq.(\ref{phi_p_plus}) we obtain, after calculating
the integrals with the delta functions, that 
\begin{equation}
\begin{split}
\frac{{\mbox d}\hspace{1ex}}{{\mbox d}x}\mathcal{R}^{2-\alpha}\psi(x) & = 
2i C_1\left(\frac{\lambda}{\hbar}\right)^{\alpha-1} {\rm e}^{i\lambda x /\hbar} 
- 2i C_2\left(\frac{\lambda}{\hbar}\right)^{\alpha-1} {\rm e}^{-i\lambda x /\hbar} \\
& \quad + \frac{\alpha}{\pi}(C_1+C_2)\Omega_\alpha \left(\frac{\lambda}{\hbar}\right)^{\alpha-1} 
\Xi_\alpha\left(\frac{\lambda x}{\hbar}\right) , 
\end{split}
\end{equation}
where 
\begin{equation}
\Xi_\alpha\left(w\right) = \int_0^\infty \sin{w q}\frac{q^{\alpha-1}}{q^\alpha -1}{\mbox d} q .
\end{equation}
This integral can be calculated in a way analogous as we did in Appendix B. The result is 
\begin{equation}
\label{int_Xi}
\begin{split}
\Xi_\alpha\left(w\right) & = 
\frac{\pi}{2}\operatorname{sign}(w) \bigg(
 H_{2,3}^{2,1} \left[ |w|^\alpha  \left| 
\begin{array}{l}
(0,1), (0,(2+\alpha)/2) \\
(0,\alpha), (0,1), (0,(2+\alpha)/2) \end{array} \right. \right] \\
& \quad - H_{2,3}^{2,1} \left[ |w|^\alpha   \left| 
\begin{array}{l}
(0,1), (0,(2-\alpha)/2) \\
(0,\alpha), (0,1), (0,(2-\alpha)/2) \end{array} \right. \right] \bigg) .   
\end{split}
\end{equation}
In order to study the limit $x\rightarrow 0^\pm$ we need to use 
\begin{equation}
\lim_{w\rightarrow 0} H_{2,3}^{2,1} \left[ |w|^\alpha  \left| 
\begin{array}{l}
(0,1), (0,\beta) \\
(0,\alpha), (0,1), (0,\beta) \end{array} \right. \right] = 
\frac{\beta}{\alpha} , 
\end{equation}
which can be calculated using the definition of the Fox's $H$-function 
and the residue theorem (the RHS comes from the residue at $w = 0$), 
and which gives  
\begin{equation}
\lim_{w\rightarrow 0^\pm} \Xi_\alpha(w) = \pm \frac{\pi}{2} . 
\end{equation}
Using this result, we have 
\begin{equation}
\lim_{x\rightarrow 0^\pm}  
\frac{{\mbox d}\hspace{1ex}}{{\mbox d}x}\mathcal{R}^{2-\alpha}\psi(x) = 
\pm \frac{\alpha}{2}(C_1+C_2)\Omega_\alpha \left(\frac{\lambda}{\hbar}\right)^{\alpha-1} . 
\end{equation}
On the other hand, $\psi(0)$ can be calculated using Eq.(\ref{EqJ0}) 
in Eq.(\ref{fourier.inv.delta.2}), which gives 
\begin{equation}
\psi(0) = \frac{(C_1+C_2)\alpha \Omega_\alpha \lambda^{\alpha-1}}{2\pi \gamma} = 
\frac{\hbar^\alpha D_\alpha}{V_0} \alpha (C_1+C_2)  
\Omega_\alpha \left(\frac{\lambda}{\hbar}\right)^{\alpha-1}  ,  
\end{equation}
where we used the definition of $\gamma$ in Eq.(\ref{def_gamma}). 
We see, therefore, that Eq.(\ref{bc1}) is satisfied.

\end{document}